\documentclass[referee]{aa} 

\usepackage{graphicx}
\usepackage{txfonts}
\bibpunct{(}{)}{;}{a}{}{,} 
\begin{document}

   \title{No variations in transit times for Qatar-1 b\thanks{Partly based on (1) data collected with telescopes at the Rozhen National Astronomical Observatory and (2) observations obtained with telescopes of the University Observatory Jena, which is operated by the Astrophysical Institute of the Friedrich-Schiller-University.}}

  
   \author{G.~Maciejewski\inst{\ref{inst1}},
          M.~Fern\'andez\inst{\ref{inst2}},
          F.~J.~Aceituno\inst{\ref{inst2}},
          J.~Ohlert\inst{\ref{inst3},\ref{inst4}},
          D.~Puchalski\inst{\ref{inst1}},
          D.~Dimitrov\inst{\ref{inst5}},
          M.~Seeliger\inst{\ref{inst6}},
          M.~Kitze\inst{\ref{inst6}},
          St.~Raetz\inst{\ref{inst6},\ref{inst7}},
          R.~Errman\inst{\ref{inst6},\ref{inst8}},
          H.~Gilbert\inst{\ref{inst6}},
          A.~Pannicke\inst{\ref{inst6}},
          J.-G.~Schmidt\inst{6},
          \and
          R.~Neuh\"auser\inst{6}
    }

   \institute{Centre for Astronomy, Faculty of Physics, Astronomy and Informatics, 
             Nicolaus Copernicus University, Grudziadzka 5, 87-100 Torun, Poland, 
              \email{gmac@umk.pl}\label{inst1}
         \and
             Instituto de Astrof\'isica de Andaluc\'ia (IAA-CSIC), Glorieta de la Astronom\'ia 3, 18008 Granada, Spain\label{inst2}
         \and
             Michael Adrian Observatorium, Astronomie Stiftung Trebur, 65428 Trebur, Germany\label{inst3}
         \and 
             University of Applied Sciences, Technische Hochschule Mittelhessen, 61169 Friedberg, Germany\label{inst4}
         \and
             Institute of Astronomy, Bulgarian Academy of Sciences, 72 Tsarigradsko Chausse Blvd., 1784 Sofia, Bulgaria\label{inst5}
         \and
             Astrophysikalisches Institut und Universit\"ats-Sternwarte, Schillerg\"asschen 2--3, 07745 Jena, Germany\label{inst6}
         \and
            European Space Agency, ESTEC, SRE-S, Keplerlaan 1, 2201 AZ Noordwijk, The Netherlands\label{inst7}
         \and
             Institute of Applied Physics, Abbe Center of Photonics, Friedrich-Schiller-Universit\"at Jena, Max-Wien-Platz 1, 07743 Jena, Germany\label{inst8}
    }


  \abstract
   {The transiting hot Jupiter planet Qatar-1~b was presented to exhibit variations in transit times that could be of perturbative nature. A hot Jupiter with a planetary companion on a nearby orbit would constitute an unprecedented planetary configuration, important for theories of formation and evolution of planetary systems. We performed a photometric follow-up campaign to confirm or refute transit timing variations. We extend the baseline of transit observations by acquiring 18 new transit light curves acquired with 0.6-2.0 m telescopes. These photometric time series, together with data available in the literature, were analyzed in a homogenous way to derive reliable transit parameters and their uncertainties. We show that the dataset of transit times is consistent with a linear ephemeris leaving no hint for any periodic variations with a range of 1 min. We find no compelling evidence for the existence of a close-in planetary companion to Qatar-1~b. This finding is in line with a paradigm that hot Jupiters are not components of compact multi-planetary systems. Based on dynamical simulations,  we place tighter constraints on a mass of any fictitious nearby planet in the system.  Furthermore, new transit light curves allowed us to redetermine system parameters with the precision better than that reported in previous studies. Our values generally agree with previous determinations. }


   \maketitle
%

\section{Introduction}
The transiting planet Qatar-1~b \citep{2011MNRAS.417..709A} is the first planetary body discovered by the Qatar Exoplanet Survey \citep{2013AcA....63..465A}. It is a typical hot Jupiter on a 1.42~d circular orbit. \citet{2011MNRAS.417..709A} give a mass of $1.09\pm0.08$~$M_{\rm{Jup}}$ (Jupiter's mass), but new radial velocity (RV) measurements by \citet{2013A&A...554A..28C} show that the planet has the greater mass of $1.33\pm0.05$~$M_{\rm{Jup}}$. The published studies provide consistent determinations of a planetary radius which is greater by 16--18\% than $R_{\rm{Jup}}$ (Jupiter's radius). The planet produces transits which are nearly 1.5 h long and 2.3\% deep in flux. The host star, GSC 4240-470 ($V=12.84$ mag), is a metal-rich dwarf of spectral type K3. It exhibits a moderate chromospheric activity, which could be partly induced by the close-in planet \citep{2013A&A...554A..28C}. The RV observations throughout a transit show a signature of the Rositter-McLaughlin effect that is consistent with a sky-projected obliquity close to zero \citep{2013A&A...554A..28C}.   

\begin{table*}
\caption{Details on instruments taking part in the campaign.} 
\label{Tab.01}      
\centering                  
\begin{tabular}{l l c l c}      
\hline\hline                
\# & Telescope       &  Diameter~(m) & CCD detector & $N_{\rm{tr}}$ \\
    & Observatory    &  & Size of matrix and field of view &  \\
\hline 
1 &  Ritchey-Chr\'etien-Coud\'e Telescope                        &  2.0 & Roper Scientific VersArray 1300B  & 1 \\
   &  National Astronomical Observatory Rozhen, Bulgaria &   & $1340 \times 1300$ pixels, $5\farcm8 \times 5\farcm6$  & \\
2 &  Ritchey-Chr\'etien Telescope                            &  1.5 & Roper Scientific VersArray 2048B  & 4 \\
   &  Sierra Nevada Observatory (OSN), Spain            &   & $2048 \times 2048$ pixels, $7\farcm92 \times 7\farcm92$  & \\
3 &  Trebur 1-meter Telescope                                &  1.2 & SBIG STL-6303  & 3 \\
   &  Michael Adrian Observatory, Trebur, Germany &   & $3072 \times 2048$ pixels, $10\farcm0 \times 6\farcm7$  & \\
4 &  Schmidt Teleskop Kamera \citep{2010AN....331..449M}  & 0.9 & E2V CCD-42-40  & 3 \\
   &  University Observatory Jena, Germany                            &    & $2048 \times 2048$ pixels, $52\farcm8 \times 52\farcm8$  & \\
5 &  Cassegrain Telescope                                       &  0.6 & SBIG STL-1001  & 6 \\
   &  Nicolaus Copernicus University, Poland           &   & $1024 \times 1024$ pixels, $11\farcm8 \times 11\farcm8$  & \\
6 &  Cassegrain Telescope                                       &  0.6 & SBIG STL-6303  & 1 \\
   &  Volkssternwarte Kirchheim, Germany              &   & $3072 \times 2048$ pixels, $52\farcm4 \times 34\farcm9$  & \\
\hline                                   
\end{tabular}
\tablefoot{Diameter is the size of a telescope's main mirror. $N_{\rm{tr}}$ is the total number of complete or partial transit light curves, acquired with the given instrument.}
\end{table*}

\begin{table*}
\caption{Details on individual light curves reported in this paper.} 
\label{Tab.02}      
\centering                  
\begin{tabular}{l l c c c c l}      
\hline\hline                
Date UT (epoch) & Telescope &  Filter  & $\Gamma$ & $pnr$ &    $X$   & Sky  conditions\\
\hline 
2011 Aug 01 (181) & Volk.Kirch. 0.6 m & none &  0.87 & 3.68 & $1.05 \rightarrow 1.03 \rightarrow 1.05$ & Clear \\
2012 Dec 13 (533) & Torun 0.6 m         & none &  1.06 & 2.43 & $1.12 \rightarrow 1.55$ & Clear \\
2013 Sep 03 (719) & Torun 0.6 m         & none &  1.51 & 2.51 & $1.02 \rightarrow 1.14$ & Clear \\
2013 Sep 30 (738) & Torun 0.6 m         & none &  1.24 & 2.09 & $1.03 \rightarrow 1.22$ & Photometric\\
2013 Nov 26 (778) & Trebur 1.2 m       & none &  1.39 & 1.18 & $1.07 \rightarrow 1.14$ & Photometric \\
2013 Dec 20 (795) & Jena 0.9 m       & $R_{\rm{B}}$ &  1.12 & 2.67 & $1.23 \rightarrow 1.81$ & Clear \\
2014 Mar 25 (862) & Trebur 1.2 m       & none &  0.40 & 4.36 & $2.10 \rightarrow 1.43$ & Passing clouds \\
2014 Apr 02 (867) & Rozhen 2.0 m  & $R_{\rm{C}}$ &  1.22 & 1.47 & $1.60 \rightarrow 1.22$ & Photometric \\
                             & Torun 0.6 m         & none &  1.82 & 2.36 & $1.41 \rightarrow 1.14$ & Clear \\
2014 May 16 (898) & OSN 1.5 m      & $R_{\rm{C}}$ &  1.09 & 2.21 & $1.48 \rightarrow 1.14$ & Mostly clear, some high clouds \\
2014 May 26 (905) & OSN 1.5 m      & $R_{\rm{C}}$ &  1.33 & 1.39 & $1.62 \rightarrow 1.16$ & Mostly photometric \\
2014 Jun 22 (924) & OSN 1.5 m      & $R_{\rm{C}}$ &  1.09 & 1.61 & $1.37 \rightarrow 1.13$ & Mostly photometric, occasionally \\
  &    &  &   &  &  & passing high clouds before ingress\\
2014 Jul 01 (931) & Trebur 1.2 m   & $R_{\rm{B}}$ &  0.88 & 2.64 & $1.17 \rightarrow 1.03$ & Photometric, first half of series \\
  &    &  &   &  &  & partly obscured by the dome slit \\
2014 Sep 10 (981) & OSN 1.5 m      & $R_{\rm{C}}$ &  1.33 & 1.29 & $1.14 \rightarrow 1.13 \rightarrow 1.37$ & Photometric \\
2014 Sep 18 (986) & Jena 0.9 m       & $R_{\rm{B}}$ &  1.16 & 2.25 & $1.19 \rightarrow 1.66$ & Photometric \\
2014 Sep 27 (993) & Jena 0.9 m       & $R_{\rm{B}}$ &  1.14 & 2.02 & $1.12 \rightarrow 1.54$ & Photometric \\
                             & Torun 0.6 m     & $R_{\rm{C}}$ &  0.95 & 3.74 & $1.14 \rightarrow 1.53$ & Clear, high humidity \\
2014 Oct 27 (1014) & Torun 0.6 m  & $R_{\rm{C}}$ &  0.86 & 3.45 & $1.05 \rightarrow 1.21$ & High humidity, occasionally passing\\ 
  &    &  &   &  &  & high clouds \\
\hline                                   
\end{tabular}
\tablefoot{Date UT is given for mid-transit time. Epoch is the transit number from the initial ephemeris given in \citet{2011MNRAS.417..709A}. $R_{\rm{C}}$ and $R_{\rm{B}}$ denote Cousins and Bessell $R$-band filters, respectively. $\Gamma$ is the median number of exposures per minute. $pnr$ is the photometric scatter in millimag per minute as defined by \citet{2011AJ....142...84F}. $X$ shows the airmass change during transit observations.}
\end{table*}

More recently, \citet{2013A&A...555A..92V} have found indications for long-term variations in transit timing residuals from a linear ephemeris. The periodicity of this signal is postulated to be close to 190 or 390~d with a peak-to-peak amplitude of about 2~min. These variations were interpreted as a result of the exchange of energy and angular momentum via gravitational interaction with an unseen, close-in planetary companion \citep{2005Sci...307.1288H,2005MNRAS.359..567A}. Such planetary configuration would be unprecedented because hot Jupiters appear to be alone or accompanied by other planetary bodies on wide orbits \citep[e.g.,][]{2012PNAS..109.7982S}. On the other hand, studies of candidates for hot planets from a Kepler sample show that some fraction of these objects reveal periodic variations in transit timing that could be dynamically induced \citep{2013A&A...553A..17S}. There may be only a small fraction of hot Jupiters which are a part of compact multiple-planet systems. If such systems are confirmed, their properties will place constraints on planet formation and migration theories. In some circumstances, migration models, based on tidal interactions between a giant planet and a circumstellar gas disk, predict the presence of neighboring low-mass planets, close to or trapped in interior or exterior mean-motion resonances (MMRs) with hot Jupiters \citep[e.g.,][]{2009MNRAS.397.1995P}. Such resonant or near-resonant companions would induce orbital perturbations producing detectable variations in transit times. The analysis of multiple-planet candidate systems shows that most planetary systems are not resonant, but the distribution of planet period ratios shows clumping just wide of low-order resonances, such as 2:1 or 3:2 \citep{2014ApJ...790..146F}. The origin of this pattern is unclear. The question whether hot Jupiters are accompanied by planets on nearby orbits or not, and what is the fraction of such planetary configurations still remains open.

In this context, Qatar-1~b with its preliminary detection of the transit time variation (TTV) signal could become a prototype of a new group of multiple-planet packed systems with hot Jupiters.  As the system deserves further investigation, we organized an observing campaign to confirm or refute perturbations in the orbital motion of Qatar-1~b by the method of transit timing.
 
 \begin{figure*}
    \centering
    \includegraphics[width=18cm]{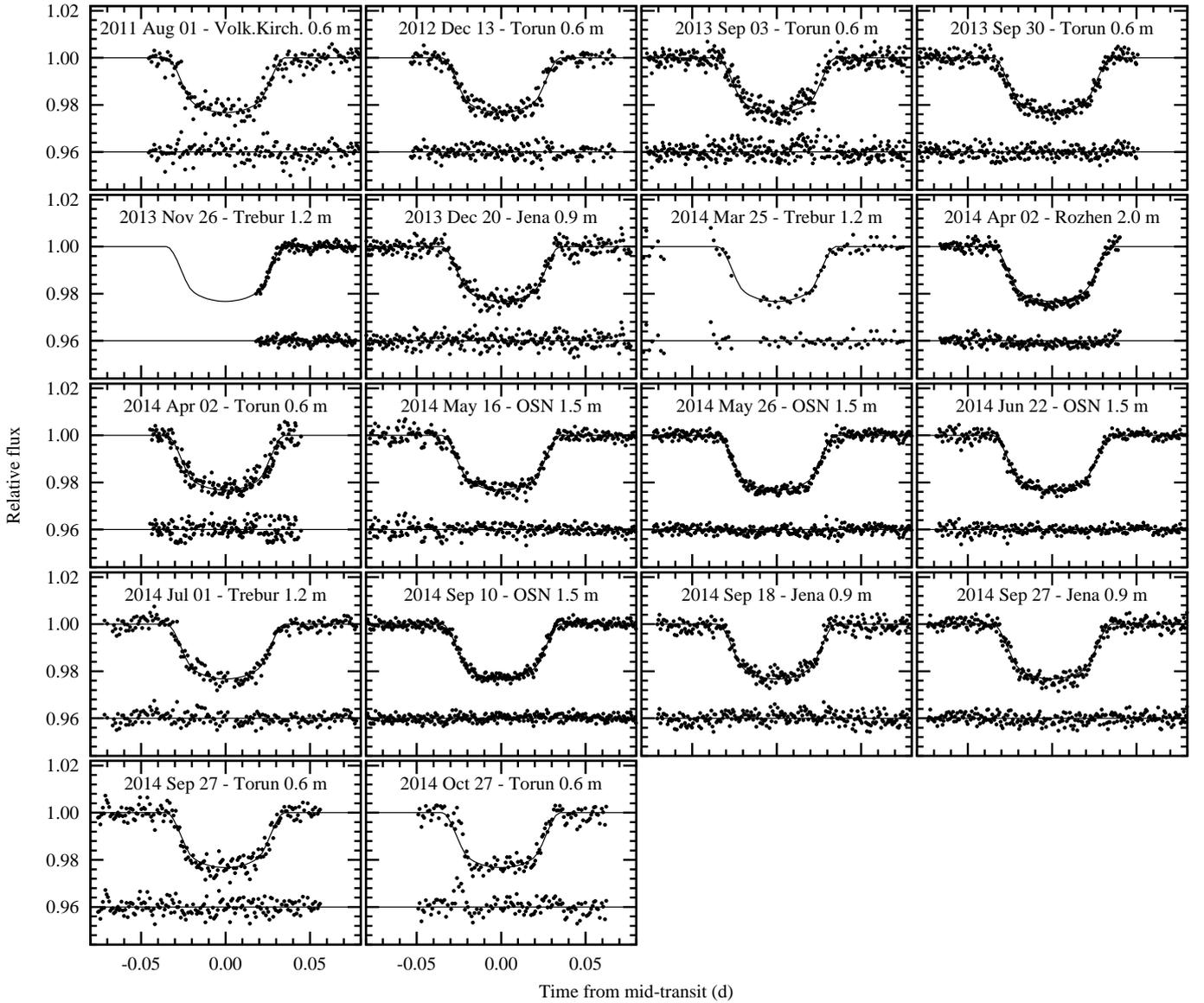}
    \caption{New transit light curves acquired for Qatar-1~b, sorted by observation dates. The continuous lines show the best-fitting model, the residuals are plotted below each light curve.}
   \label{Fig.LCs}%
  \end{figure*}

\section{Observations}

We acquired 18 new light curves for 16 transit events with six 0.6-2.0 m telescopes located across Europe (in Bulgaria, Poland, Germany, and Spain). The characteristics of individual instruments and detectors are given in Table~\ref{Tab.01}. Two transits were observed simultaneously from two distinct sites.  Most of the observations were performed in $R$ filter, in which the instrumental setups were found to be the most effective. We also acquired light curves in white light, i.e.\ without any filter, to increase the signal-to-noise ratio and obtain more precise data for timing purposes. Telescopes were moderately defocused to observe the Qatar-1 star with exposure times between 10 and 60 s, depending on mirror diameter, filter, and observing conditions. For a given light curve, the exposure time was kept fixed throughout a run. The timestamps were saved in coordinated universal time (UTC), provided by a GPS system or the network time protocol software. Details on individual runs are presented in Table~\ref{Tab.02}. 

Photometric time series were obtained with the differential aperture photometry applied to science images after bias and dark subtraction followed by flat-field correction with sky flats. The radius of the photometric aperture was optimized for each dataset to achieve the smallest scatter in the out-of-transit light curves. Magnitudes were determined against available comparison stars. Each light curve was de-trended by fitting a parabola along with a trial transit model, then converted to fluxes and normalized to unity outside of the transit event. The timestamps were converted to barycentric Julian dates in barycentric dynamical time \citep[$\rm{BJD_{TDB}}$,][]{2010PASP..122..935E}.

\section{Results}

\subsection{System parameters}

The Transit Analysis Package \citep[TAP,][]{2012AdAst2012E..30G} was used to model the set of 18 new transit light curves, supplemented with 31 complete or partial light curves available in the literature. \citet{2013A&A...554A..28C} provide 5 high-quality photometric time series acquired between May 2011 and September 2012 with the 1.82~m Asiago (Italy) and 1.23 m Calar Alto (Spain) telescopes. \citet{2013A&A...555A..92V} make available 26 light curves. They were obtained from March 2011 to August 2012 with the 1.2 m Oskar-L\"uhning telescope at Hamburg Observatory (Germany) and the 0.6 m Planet Transit Study Telescope at the Mallorca Observatory (Spain). The TAP code uses the Markov Chain Monte Carlo (MCMC) method with the Metropolis-Hastings algorithm and a Gibbs sampler to find the best-fit transit model, based on the analytical approach of \citet{2002ApJ...580L.171M}. As the photometric time series may be afflicted by time-correlated noise, the wavelet-based algorithm of \citet{2009ApJ...704...51C} is employed to robustly estimate parameter uncertainties. A quadratic limb darkening (LD) law \citep{1950HarCi.454....1K} is implemented to approximate the flux distribution across the stellar disk. The values of linear and quadratic LD coefficients, $u_1$ and $u_2$, respectively, were linearly interpolated from tables of \citet{2011A&A...529A..75C} using a dedicated tool of the EXOFAST package \citep{2013PASP..125...83E}. The stellar parameters, such as effective temperature $T_{\rm{eff}} = 4910 \pm 100$ K, surface gravity $\log g = 4.55 \pm 0.1$ (in cgs units), and metallicity $\rm{[Fe/H]} = 0.2 \pm 0.1$, were taken from \citet{2013A&A...554A..28C}. The efficient maximum of the energy distribution in white light was found to coincide with the $R$ band, so $R$-band LD coefficients were adopted for white light curves.

\begin{table*}
\caption{Redetermined system parameters, together with values from the literature.} 
\label{Tab.03}      
\centering                  
\begin{tabular}{l c c c c}      
\hline\hline                
Parameter & This paper &  Alsubai et al.\ (2011)\tablefootmark{a}  & Covino et al.\ (2013) & von Essen et al.\ (2013) \\
\hline 
Orbital inclination, $i_{\rm{b}}$ (deg) & $84.26^{+0.17}_{-0.16}$ & $83.47^{+0.40}_{-0.36}$ & $83.82 \pm 0.25$ & $84.52 \pm 0.24$\\
Scaled semi-major axis, $a_{\rm{b}}/R_{*}$ & $6.319^{+0.070}_{-0.068}$ & --- & $6.25 \pm 0.10$ & $6.42 \pm 0.10$\\
Planet to star radii ratio, $R_{\rm{b}}/R_{*}$ & $0.14591^{+0.00076}_{-0.00078}$ & $0.1453 \pm 0.0016$ & $0.1513 \pm 0.0008$ & $0.1435 \pm 0.0008$\\
Transit parameter, $b=\frac{a_{\rm{b}}}{R_{*}}\cos{i_{\rm{b}}}$ & $0.63 \pm 0.02$ & $0.696^{+0.021}_{-0.024}$ & $0.675 \pm 0.016$ & ---\\
Transit total duration, $T_{14}$ (min) & $98.5 \pm 1.7$ & --- & $97.6 \pm 1.4$ & ---\\
Semi-major axis, $a_{\rm{b}}$ (AU)  & $0.02298^{+0.00069}_{-0.00036}$ & $0.02343^{+0.00026}_{-0.00025}$ & $0.0234 \pm 0.0012$ & ---\\
Planetary mass\tablefootmark{b}, $M_{\rm{b}}$ ($M_{\rm{Jup}}$) & $1.275^{+0.079}_{-0.043}$ & $1.090^{+0.084}_{-0.081}$ & $1.33 \pm 0.05$ & ---\\
Planetary radius, $R_{\rm{b}}$ ($R_{\rm{Jup}}$) & $1.136^{+0.037}_{-0.022}$ & $1.164 \pm 0.045$ & $1.18 \pm 0.09$ & ---\\
Planetary density, $\rho_{\rm{b}}$ ($\rho_{\rm{Jup}}$) & $0.87^{+0.10}_{-0.06}$ & $0.690^{+0.098}_{-0.084}$ & $0.80 \pm 0.20$ & ---\\
Planetary gravity\tablefootmark{b}, $g_{\rm{b}}$ (m~s$^{-2}$) & $25.65 \pm 0.71$ & $18.4^{+2.0}_{-1.8}$ & $23.6^{+1.4}_{-1.3}$ & ---\\
Stellar mass, $M_{*}$  ($M_{\odot}$) & $0.803^{+0.072}_{-0.038}$ & $0.85\pm 0.03$ & $0.85 \pm 0.03$ & ---\\
Stellar radius, $R_{*}$  ($R_{\odot}$)  & $0.782^{+0.025}_{-0.015}$ & $0.823 \pm 0.025$ & $0.80 \pm 0.05$ & ---\\
Stellar density, $\rho_{*}$ ($\rho_{\odot}$) & $1.680^{+0.056}_{-0.055}$ & $1.52 \pm 0.12$ & $1.62 \pm 0.08$ & ---\\
Stellar gravity\tablefootmark{c}, $\log g_{*}$ (cgs) & $4.556^{+0.020}_{-0.016}$ & $4.536 \pm 0.024$ & $4.55 \pm 0.10$ & ---\\
Stellar luminosity, $L_{*}$  ($L_{\odot}$) & $0.318^{+0.051}_{-0.025}$ & --- & --- & ---\\
Age, (Gyr) & $9.8^{+3.4}_{-5.8}$ & $>4$ & $\approx 4.5$ & ---\\
 & $$ & $$ & $$ & $$\\
\hline                                   
\end{tabular}
\tablefoot{
   \tablefoottext{a}{Values for a solution with a circular orbit were taken.} 
   \tablefoottext{b}{In calculations, the value of the RV semi-amplitude was taken from Covino et al.\ (2013).} 
   \tablefoottext{c}{Based on parameters derived from transit photometry.} 
}
\end{table*}

In a final iteration of the fitting procedure, the orbital inclination $i_{\rm{b}}$, the semimajor-axis scaled by stellar radius $a_{\rm{b}}/R_{*}$, and the planetary to stellar radii ratio $R_{\rm{b}}/R_{*}$ were linked for all light curves and fitted simultaneously. The mid-transit times were determined independently for each epoch. The orbital period $P_{\rm{b}}$ was kept fixed at a value refined in Sect.~\ref{Sect.TTimes}. The LD coefficients were allowed to vary around the theoretical values under the Gaussian penalty of $\sigma = 0.05$ to account for uncertainties in stellar parameters and possible systematic errors of the theoretical predictions \citep[e.g.,][]{2013A&A...560A.112M}. The orbit of the planet was assumed to be circular \citep{2013A&A...554A..28C}. Although the individual light curves are de-trended, the TAP code was allowed to take possible linear trends into account and include their uncertainties in the total error budget of the fit. We used ten MCMC chains, each containing $10^6$ steps. The first 10\% of the results were discarded from each chain to minimize the influence of the initial values of the parameters. The best-fitting parameters, determined as the median values of marginalized posteriori probability distributions, and 1\,$\sigma$ uncertainty estimates, defined by a range of 68.3\% values of the distributions, are given in Table~\ref{Tab.03}. The new light curves are plotted in Fig.~\ref{Fig.LCs}, together with the transit model and residuals. 

 \begin{figure}
    \centering
    \includegraphics[width=0.5\hsize]{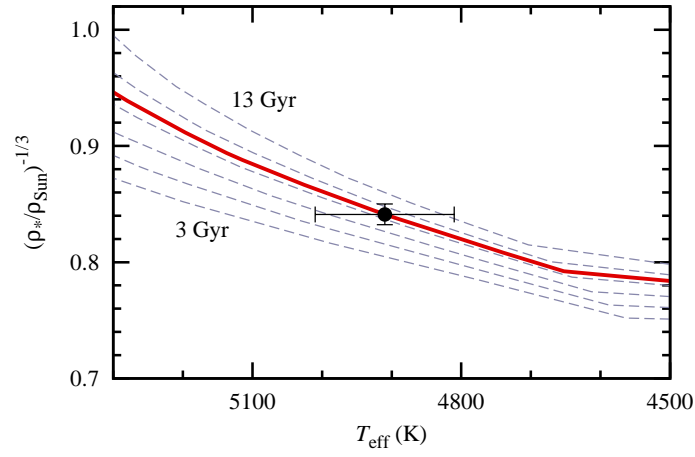}
    \caption{Modified Hertzsprung-Russel diagram with Qatar-1 marked as a central dot. The {\sc PARSEC} isochrones of $\rm{[Fe/H]}=0.20$ and ages between 3 and 13 Gyr with a step of 2 Gyr are sketched with dashed lines. The best-fitting isochrone is drawn with a continuous line.}
   \label{Fig.HR}%
  \end{figure}

The parameters derived from the transit model, combined with literature quantities determined from spectroscopic data, allowed us to redetermine the stellar mass $M_*$, luminosity $L_*$, and system age. The mean stellar density $\rho_*$ was calculated from transit parameters, independent of theoretical stellar models, using the formula
\begin{equation}
     \rho_{*} = \frac{3\pi}{G P_{\rm{b}}^2} \left(\frac{a_{\rm{b}}}{R_{*}}\right)^3\, , \; 
\end{equation}
where $G$ is the universal gravitational constant. Then, the host star was placed at a modified Hertzsprung-Russel diagram (Fig.~\ref{Fig.HR}) displaying the $\rho_{*} ^{-1/3}$ versus $T_{\rm{eff}}$ plane, together with {\sc PARSEC} isochrones in version 1.2S \citep{2012MNRAS.427..127B}. The stellar parameters and their uncertainties were derived by interpolating the sets of isochrones with $\rm{[Fe/H]}$ ranging from 0.1 to 0.3.

 \begin{figure}
    \centering
   \includegraphics[width=0.5\hsize]{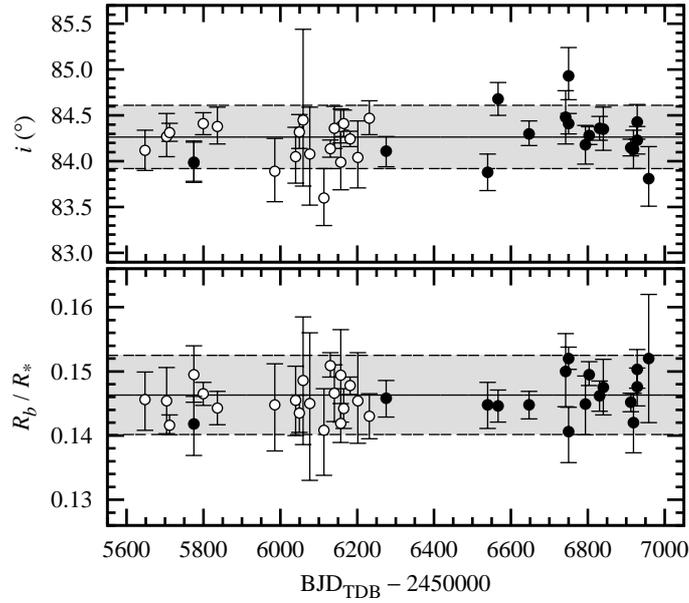}
      \caption{Distribution of $i_{\rm{b}}$ and $R_{\rm{b}}/R_{*}$ in a function of time for a trial fit, in which both parameters were allowed to vary for individual complete transit light curves. Open and filed symbols represent literature and our new datasets, respectively. Continuous lines mark weighted mean values. Grayed areas between dashed lines show uncertainties at the confidence level of 95.5\%, i.e. $2\sigma$, where $\sigma$ is the weighted standard deviation.}
         \label{Fig.Pars}
 \end{figure}

The parameters $i_{\rm{b}}$ and $R_{\rm{b}}/R_{*}$ were checked for long timescale trends or periodic variability ($a_{\rm{b}}/R_{*}$ was not considered because it is correlated with $i_{\rm{b}}$). The fitting procedure was repeated for a subset of 37 complete light curves, with $i_{\rm{b}}$ and $R_{\rm{b}}/R_{*}$ allowed to be determined independently for each dataset. The parameter $a_{\rm{b}}/R_{*}$ was allowed to vary around a best-fit value under the Gaussian penalty, as given in Table~\ref{Tab.03}. Individual determinations are plotted in Fig.~\ref{Fig.Pars}. A model assuming a constant $i_{\rm{b}}$ has a reduced chi-squared $\chi^2_{\rm{red}}$ of 1.15 and a $p$-value of 0.25. The parameter $R_{\rm{b}}/R_{*}$ is also constant with $\chi^2_{\rm{red}} = 1.24$ that corresponds to a $p$-value of 0.15. For both parameters, there is no reason to reject a null hypothesis assuming constancy. In addition, both parameters were searched for periodic variations with the Lomb-Scargle algorithm \citep{1976Ap&SS..39..447L,1982ApJ...263..835S} and no significant signal was found. The lack of any variations justifies linking these parameters in the final model.

\subsection{Transit timing}\label{Sect.TTimes}

 \begin{figure*}
    \centering
    \includegraphics[width=18cm]{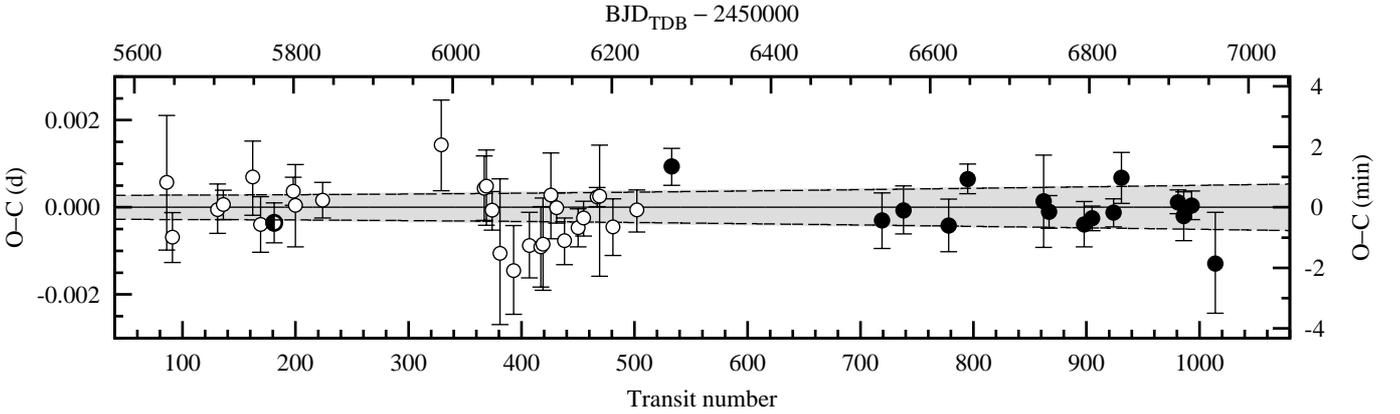}
    \caption{Residuals of transit times from the refined linear ephemeris. Open symbols mark literature data. Filled dots denote new transits reported in this paper. The mid-transit time at epoch 181 (2011 Aug 01 = BJD 2455775.4), marked with a half-filled symbol, is determined using a light curve from \citet{2013A&A...555A..92V} and a light curve reported in this paper. The grayed area between dashed lines shows the propagation of uncertainties of the ephemeris at a 95.5\% confidence level.}
   \label{Fig.TTres}%
  \end{figure*}

New mid-transit times, together with redetermined values from the literature data, allowed us to refine the transit ephemeris. As a result of a least-square linear fit, in which timing uncertainties were taken as weights, we derived 
\begin{equation}
 P_{\rm{b}}=1.42002406 \pm 0.00000021~\rm{d}
\end{equation}
and the mid-transit time at cycle zero
\begin{equation}
  T_{0}=2455518.41117 \pm 0.00014 ~\rm{BJD_{TDB}}\, . \;
\end{equation} 
The value of $\chi^2_{\rm{red}}$  was found to be equal to 1.07, clearly showing that there is no hint for any deviation from the constant orbital period. The procedure was repeated for a limited sample of 17 mid-transit times with uncertainties smaller than 40 s, hence the most reliable. Again, the linear ephemeris was found to reproduce observed transits perfectly with $\chi^2_{\rm{red}} =0.91$. This shows that any possible TTV signal is not masked  by a scatter of data points of lower quality and hence lower timing precision. Individual transit times and residuals are collected in Table~\ref{Tab.04}, and residuals from the ephemeris are plotted in Fig.~\ref{Fig.TTres}.  

\begin{figure}
    \centering
   \includegraphics[width=0.5\hsize]{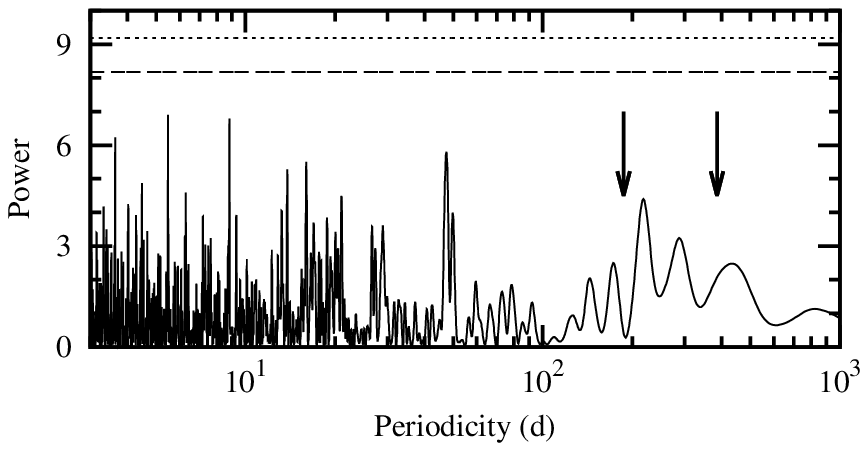}
      \caption{Lomb-Scargle periodogram for timing residuals plotted in Fig.~\ref{Fig.TTres}. Dashed and dotted horizontal lines show the 5\% and 1\% FAP levels, respectively. Arrows indicate positions of the TTV signals claimed by \citet{2013A&A...555A..92V}.}
         \label{Fig.TTper}
 \end{figure}

The Lomb-Scargle periodogram (Fig.~\ref{Fig.TTper}) reveals no significant power at periods of $187 \pm 17$ and  $386 \pm 54$ d, reported by \citet{2013A&A...555A..92V}. A bootstrap resampling method, based on $10^5$ trials of the randomly permuted timing residuals at the original observing epochs, was used to determine a false alarm probability (FAP) of 28\% for the strongest peak.

\begin{table}
\caption{New and redetermined mid-transit times.} 
\label{Tab.04}      
\centering                  
\begin{tabular}{r l l c}      
\hline\hline                
Epoch & $T_{\rm{mid}}$ (d) &  O--C (d)  & Data source \\
\hline 
  86 &   $5640.5338^{+0.0015}_{-0.0016}$  & $+0.0006$  &  1 \\
  91 & $5647.63267^{+0.00057}_{-0.00058}$ & $-0.00069$ &  1 \\
 131 & $5704.43426^{+0.00059}_{-0.00054}$ & $-0.00005$ & 1  \\
 136 & $5711.53450^{+0.00033}_{-0.00035}$ & $+0.00006$ & 1, 2  \\
 162 & $5748.45576^{+0.00083}_{-0.00088}$ & $+0.00070$ & 1  \\
 169 & $5758.39484^{+0.00064}_{-0.00064}$ & $-0.00039$ & 1  \\
 181 & $5775.43517^{+0.00046}_{-0.00046}$ & $-0.00036$ &  1, 3 \\
 198 & $5799.57630^{+0.00032}_{-0.00032}$ & $+0.00037$ &  2 \\
 200 & $5802.41602^{+0.00094}_{-0.00095}$ & $+0.00004$ &  1 \\
 224 & $5836.49672^{+0.00041}_{-0.00041}$ & $+0.00017$ &  1 \\
 329 &   $5985.6005^{+0.0010}_{-0.0011}$  & $+0.0014$  &  1 \\
 367 & $6039.56043^{+0.00075}_{-0.00073}$ & $+0.00043$ &  1 \\
 369 & $6042.40053^{+0.00083}_{-0.00090}$ & $+0.00049$ &  1 \\
 374 & $6049.50010^{+0.00043}_{-0.00046}$ & $-0.00007$ &  1 \\
 381 &   $6059.4393^{+0.0017}_{-0.0016}$  & $-0.0011$  &  1 \\
 393 &   $6076.4792^{+0.0010}_{-0.0010}$  & $-0.0015$  &  1 \\
 407 & $6096.36008^{+0.00076}_{-0.00075}$ & $-0.00087$ & 1  \\
 417 & $6110.56029^{+0.00092}_{-0.00092}$ & $-0.00091$ & 1  \\
 419 &   $6113.4004^{+0.0011}_{-0.0011}$  & $-0.0009$  &  1 \\
 426 & $6123.34169^{+0.00097}_{-0.00100}$ & $+0.00028$ & 1  \\
 431 & $6130.44153^{+0.00034}_{-0.00035}$ & $-0.00001$ &  1, 2 \\
 438 & $6140.38094^{+0.00052}_{-0.00055}$ & $-0.00077$ &  1 \\
 450 & $6157.42152^{+0.00043}_{-0.00044}$ & $-0.00047$ &  1 \\
 455 & $6164.52187^{+0.00038}_{-0.00042}$ & $-0.00024$ &  2 \\
 467 & $6181.56264^{+0.00022}_{-0.00023}$ & $+0.00024$ &  2 \\
 469 &   $6184.4027^{+0.0012}_{-0.0018}$  & $+0.0003$  & 1  \\
 481 & $6201.44229^{+0.00064}_{-0.00066}$ & $-0.00045$ & 1  \\
 502 & $6231.26318^{+0.00046}_{-0.00051}$ & $-0.00006$ & 1  \\
 533 & $6275.28493^{+0.00041}_{-0.00043}$ & $+0.00094$ & 3 \\
 719 & $6539.40816^{+0.00064}_{-0.00065}$ & $-0.00030$ & 3 \\
 738 & $6566.38885^{+0.00057}_{-0.00054}$ & $-0.00007$ & 3 \\
 778 & $6623.18947^{+0.00060}_{-0.00060}$ & $-0.00042$ & 3 \\
 795 & $6647.33094^{+0.00035}_{-0.00033}$ & $+0.00065$ & 3 \\
 862 &   $6742.4720^{+0.0011}_{-0.0011}$  & $+0.00014$ & 3 \\
 867 & $6749.57192^{+0.00038}_{-0.00037}$ & $-0.00011$ & 3 \\
 898 & $6793.59237^{+0.00053}_{-0.00051}$ & $-0.00040$ & 3 \\
 905 & $6803.53269^{+0.00028}_{-0.00028}$ & $-0.00025$ & 3 \\
 924 & $6830.51327^{+0.00032}_{-0.00033}$ & $-0.00012$ & 3 \\
 931 & $6840.45424^{+0.00058}_{-0.00058}$ & $+0.00068$ & 3 \\
 981 & $6911.45488^{+0.00029}_{-0.00026}$ & $+0.00012$ & 3 \\
 986 & $6918.55468^{+0.00055}_{-0.00056}$ & $-0.00020$ & 3 \\
 993 & $6928.49510^{+0.00033}_{-0.00032}$ & $+0.00004$ & 3 \\
1014 &   $6958.3143^{+0.0012}_{-0.0011}$  & $-0.0013$  & 3 \\
\hline                                   
\end{tabular}
\tablefoot{Epoch is the transit number from the initial ephemeris given in \citet{2011MNRAS.417..709A}, $T_{\rm{mid}}$ is the mid-transit time in BJD$_{\rm{TDB}}$ minus 2450000 d. O--C is the value of timing residuals from the refined transit ephemeris. Data source: (1)~\citet{2013A&A...555A..92V}; (2)~\citet{2013A&A...554A..28C}; (3)~this paper.
}
\end{table}

\subsection{Constraints on an additional planet}\label{Sect.Constraints}

The TTV method is particularly sensitive to perturbers close to MMRs. The non-detection of periodic variations in transit timing allows such planetary configurations to be ruled out. In remaining configurations, planetary companions may be detected easier in the domain of precise Doppler measurements. Combining these two methods together places constraints on an upper mass of any hypothetical second planet in the system as a function of its semi-major axis. 

We followed the method which we applied to WASP-3 and WASP-1 systems \citep{2013AJ....146..147M,2014AcA....64...27M}. The Bulirsch--Stoer integrator, implemented in the Mercury 6 code \citep{1999MNRAS.304..793C}, was used to predict deviations from a Keplerian motion of Qatar-1~b caused by a fictitious perturbing planet. The system was assumed to be coplanar, with both orbits initially circular. The mass of the perturber was set at 0.5, 1, 5, 10, 50, 100, and 500 $M_{\rm{Earth}}$ (Earth masses), and the initial semi-major axis varied between 0.023 and 0.107 AU (corresponding to the orbital period ratio between 1 and 10) with a step of $2 \times 10^{-6}$ AU. Only configurations with outer perturber were considered because of the short orbital period of Qatar-1~b. The initial orbital longitude of the transiting planet was set at a value calculated for cycle zero, and the initial longitude of the fictitious perturber was shifted by $180^{\circ}$. The calculations covered a time span of 1500 days in which transits of Qatar-1~b were observed. The residuals from a linear ephemeris for synthetic datasets were compared with the observed $rms$ of transit timing, and a TTV upper mass of a fictitious planet was calculated. 

 \begin{figure}
       \centering
   \includegraphics[width=0.5\hsize]{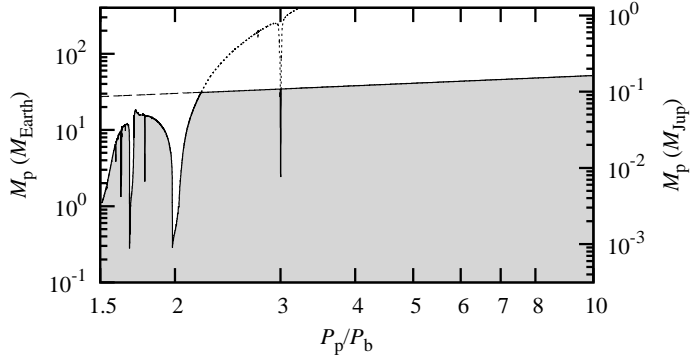}
      \caption{Upper mass limit for a fictitious outer planet in the planetary system around Qatar-1, based on timing and RV datasets, as a function of the orbital period of that planet, $P_{\rm{p}}$. Configurations that are below the detection threshold of both techniques are placed in the grayed space of parameters, bounded by a continuous line. The dotted and dashed lines show detection limits for timing and RV techniques, respectively. Configurations with $P_{\rm{p}}/P_{\rm{b}}<1.5$ were found to be unstable in a short timescale of the simulations.}
         \label{Fig.Masslimit}
 \end{figure}

Re-analysis of the RV dataset from \citet{2013A&A...554A..28C}, performed with the Systemic 2.16 \citep{2009PASP..121.1016M}, gives a circular single-planet orbit with $rms_{\rm{RV}} = 11.1$~m~s$^{-1}$. The RV measurements were acquired with the High Accuracy Radial velocity Planet Searcher-North (HARPS-N) spectrograph installed at the 3.58 m Telescopio Nazionale Galileo (TNG) on La Palma (Spain). The RV uncertainties are between 4.4 and 20~m~s$^{-1}$. The dataset from \citet{2011MNRAS.417..709A}, acquired with the Tillinghast Reflector Echelle Spectrograph (TRES) coupled with the 1.5-m telescope at the Fred L.\,Whipple Observatory (USA), was omitted because of the much lower precision of the RV measurements ($22-61$~m~s$^{-1}$). The value of $rms_{\rm{RV}}$ was transformed into the RV mass limit of the hypothetical second planet on a circular orbit as a function of its semi-major axis. 

Finally, the TTV and RV criteria were combined together to find planetary configurations with a hypothetical second planet below the detection threshold. Results are illustrated in Fig.~\ref{Fig.Masslimit}. The TTV technique is more sensitive for configurations within the 2:1 resonance. We can eliminate perturbers with masses between 10 and 20 $M_{\rm{Earth}}$ out of MMRs, and probe for a sub-Earth mass regime in 2:1 and 5:3 MMRs. For orbital periods longer than about 2.2 $P_{\rm{b}}$, the Doppler dataset provides tighter constraints except a 3:1 MMR, in which a perturber with a mass down to about 3 $M_{\rm{Earth}}$ can be eliminated thanks to the TTV method.

\section{Concluding discussion}

Despite the conservative approach, our light curve modeling provides system parameters with the smallest uncertainties published so far. This is the result of the homogenous analysis of the rich set of 18 new and 31 literature light curves. The orbital inclination $i_{\rm{b}}=84\fdg26^{+0\fdg17}_{-0\fdg16}$ agrees perfectly with the value of $84\fdg52 \pm 0\fdg24$ reported by \citet{2013A&A...555A..92V}, and differs by 1.1 and 1.4 $\sigma$ from results of \citet{2013A&A...554A..28C} and \citet{2011MNRAS.417..709A}, respectively. The scaled semi-major axis $a_{\rm{b}}/R_{*}=6.319^{+0.070}_{-0.068}$ is between $6.25 \pm 0.10$ and $6.42 \pm 0.10$ brought by \citet{2013A&A...554A..28C} and \citet{2013A&A...555A..92V}, respectively. The planet to star radii ratio $R_{\rm{b}}/R_{*}=0.14591^{+0.00076}_{-0.00078}$ is consistent with $0.1453 \pm 0.0016$ derived by \citet{2011MNRAS.417..709A}, and differs by 1.5 and 3.5 $\sigma$ from values reported by \citet{2013A&A...555A..92V} and \citet{2013A&A...554A..28C}, respectively. Planetary physical properties derived in this study confirm results of \citet{2013A&A...554A..28C}, and deviate slightly from values reported by \citet{2011MNRAS.417..709A} due to the planetary mass underestimated  in that study. Stellar parameters were found to agree with values determined in previous studies. The system's age is $\approx 10$ Gyr with the lower uncertainty of $\approx 6$ Gyr and the upper one limited by the age of the Universe. The host star is in the middle of its lifetime on the main sequence that is estimated to be $\approx 18$ Gy for a 0.8 $M_{\odot}$ star. 

Spectral observations of the Ca II H and K lines show that the host star exhibits a  moderate chromospheric activity with $\log R'_{\rm{HK}} = -4.60$ \citep{2013A&A...554A..28C}. However, we find no statistically significant variation in planet to star radii ratio that could be caused by the evolving distribution of dark spots or bright faculae outside the projected path of the planet on the stellar disk \citep[e.g.,][]{2009A&A...505.1277C}. We also identified no features in the transit light curves which could be attributed to starspot occultations by the planetary disk. 

Transit timing observations clearly show that the orbital motion of Qatar-1~b is not perturbed by any body which could produce periodic deviations with a range greater than 1 minute. We find no evidence which confirms the TTV signal with the periodicity and amplitude reported by \citet{2013A&A...555A..92V}. Combining the transit timing and RV datasets, we can rule out three proposed scenarios with perturbers in 2:1, 5:2, and 3:1 resonances even for circular orbits (in general, eccentric orbits make perturbations more pronounced). The lack of the TTV signal also makes a proposed massive perturber on a wide orbit unlikely. We note, however, that further RV measurements, acquired in a course of a few hundred days, would unequivocally reject this scenario.

We conclude that Qatar-1~b has no detectable planetary companions on nearby orbits. This finding is in line with results based on a sample of hot Jupiters observed with the space-borne facilities or ground based telescopes. As the formation of hot Jupiters is not yet fully understood \citep[e.g.,][]{2012PNAS..109.7982S}, the loneliness of  Qatar-1~b  speaks in favor of inward migration theories of massive outer planets through planet-planet scattering caused by mutual dynamical perturbations \citep{1996Natur.384..619W,1996Sci...274..954R}.

\begin{acknowledgements}
MK thanks the German national science foundation Deutsche Forschungsgemeinschaft (DFG) in projects NE515/34-1.34-2 and RE 882/12-2 for financial support. SR is currently a Research Fellow at ESA/ESTEC. SR would like to thank DFG for support in the Priority Programme SPP 1385 on the ''First Ten Million Years of the Solar System'' in projects NE 515/33-1 and -2. RE thanks the Abbe-School of Photonics for support. AP acknowledges support from DFG in SFB TR7 project C2. JGS thanks DFG in SFB TR7 project B9 for support. We would like to acknowledge financial support from the Thuringian government (B 515-07010) for the STK CCD camera used in this project.
\end{acknowledgements}



\bibliographystyle{aa} 
\bibliography{qatar1} 

\begin{thebibliography}{30}
\expandafter\ifx\csname natexlab\endcsname\relax\def\natexlab#1{#1}\fi

\bibitem[{{Agol} {et~al.}(2005){Agol}, {Steffen}, {Sari}, \&
  {Clarkson}}]{2005MNRAS.359..567A}
{Agol}, E., {Steffen}, J., {Sari}, R., \& {Clarkson}, W. 2005, \mnras, 359, 567

\bibitem[{{Alsubai} {et~al.}(2013){Alsubai}, {Parley}, {Bramich}, {Horne},
  {Collier Cameron}, {West}, {Sorensen}, {Pollacco}, {Smith}, \&
  {Fors}}]{2013AcA....63..465A}
{Alsubai}, K.~A., {Parley}, N.~R., {Bramich}, D.~M., {et~al.} 2013, \actaa, 63,
  465

\bibitem[{{Alsubai} {et~al.}(2011){Alsubai}, {Parley}, {Bramich}, {West},
  {Sorensen}, {Collier Cameron}, {Latham}, {Horne}, {Anderson}, {Bakos},
  {Brown}, {Buchhave}, {Esquerdo}, {Everett}, {F{\.z}r{\'e}sz}, {Hartman},
  {Hellier}, {Miller}, {Pollacco}, {Quinn}, {Smith}, {Stefanik}, \&
  {Szentgyorgyi}}]{2011MNRAS.417..709A}
{Alsubai}, K.~A., {Parley}, N.~R., {Bramich}, D.~M., {et~al.} 2011, \mnras,
  417, 709

\bibitem[{{Bressan} {et~al.}(2012){Bressan}, {Marigo}, {Girardi}, {Salasnich},
  {Dal Cero}, {Rubele}, \& {Nanni}}]{2012MNRAS.427..127B}
{Bressan}, A., {Marigo}, P., {Girardi}, L., {et~al.} 2012, \mnras, 427, 127

\bibitem[{{Carter} \& {Winn}(2009)}]{2009ApJ...704...51C}
{Carter}, J.~A. \& {Winn}, J.~N. 2009, \apj, 704, 51

\bibitem[{{Chambers}(1999)}]{1999MNRAS.304..793C}
{Chambers}, J.~E. 1999, \mnras, 304, 793

\bibitem[{{Claret} \& {Bloemen}(2011)}]{2011A&A...529A..75C}
{Claret}, A. \& {Bloemen}, S. 2011, \aap, 529, A75

\bibitem[{{Covino} {et~al.}(2013){Covino}, {Esposito}, {Barbieri}, {Mancini},
  {Nascimbeni}, {Claudi}, {Desidera}, {Gratton}, {Lanza}, {Sozzetti}, {Biazzo},
  {Affer}, {Gandolfi}, {Munari}, {Pagano}, {Bonomo}, {Collier Cameron},
  {H{\'e}brard}, {Maggio}, {Messina}, {Micela}, {Molinari}, {Pepe}, {Piotto},
  {Ribas}, {Santos}, {Southworth}, {Shkolnik}, {Triaud}, {Bedin}, {Benatti},
  {Boccato}, {Bonavita}, {Borsa}, {Borsato}, {Brown}, {Carolo}, {Ciceri},
  {Cosentino}, {Damasso}, {Faedi}, {Mart{\'{\i}}nez Fiorenzano}, {Latham},
  {Lovis}, {Mordasini}, {Nikolov}, {Poretti}, {Rainer}, {Rebolo L{\'o}pez},
  {Scandariato}, {Silvotti}, {Smareglia}, {Alcal{\'a}}, {Cunial}, {Di
  Fabrizio}, {Di Mauro}, {Giacobbe}, {Granata}, {Harutyunyan}, {Knapic},
  {Lattanzi}, {Leto}, {Lodato}, {Malavolta}, {Marzari}, {Molinaro},
  {Nardiello}, {Pedani}, {Prisinzano}, \& {Turrini}}]{2013A&A...554A..28C}
{Covino}, E., {Esposito}, M., {Barbieri}, M., {et~al.} 2013, \aap, 554, A28

\bibitem[{{Czesla} {et~al.}(2009){Czesla}, {Huber}, {Wolter}, {Schr{\"o}ter},
  \& {Schmitt}}]{2009A&A...505.1277C}
{Czesla}, S., {Huber}, K.~F., {Wolter}, U., {Schr{\"o}ter}, S., \& {Schmitt},
  J.~H.~M.~M. 2009, \aap, 505, 1277

\bibitem[{{Eastman} {et~al.}(2013){Eastman}, {Gaudi}, \&
  {Agol}}]{2013PASP..125...83E}
{Eastman}, J., {Gaudi}, B.~S., \& {Agol}, E. 2013, \pasp, 125, 83

\bibitem[{{Eastman} {et~al.}(2010){Eastman}, {Siverd}, \&
  {Gaudi}}]{2010PASP..122..935E}
{Eastman}, J., {Siverd}, R., \& {Gaudi}, B.~S. 2010, \pasp, 122, 935

\bibitem[{{Fabrycky} {et~al.}(2014){Fabrycky}, {Lissauer}, {Ragozzine}, {Rowe},
  {Steffen}, {Agol}, {Barclay}, {Batalha}, {Borucki}, {Ciardi}, {Ford},
  {Gautier}, {Geary}, {Holman}, {Jenkins}, {Li}, {Morehead}, {Morris},
  {Shporer}, {Smith}, {Still}, \& {Van Cleve}}]{2014ApJ...790..146F}
{Fabrycky}, D.~C., {Lissauer}, J.~J., {Ragozzine}, D., {et~al.} 2014, \apj,
  790, 146

\bibitem[{{Fulton} {et~al.}(2011){Fulton}, {Shporer}, {Winn}, {Holman},
  {P{\'a}l}, \& {Gazak}}]{2011AJ....142...84F}
{Fulton}, B.~J., {Shporer}, A., {Winn}, J.~N., {et~al.} 2011, \aj, 142, 84

\bibitem[{{Gazak} {et~al.}(2012){Gazak}, {Johnson}, {Tonry}, {Dragomir},
  {Eastman}, {Mann}, \& {Agol}}]{2012AdAst2012E..30G}
{Gazak}, J.~Z., {Johnson}, J.~A., {Tonry}, J., {et~al.} 2012, Advances in
  Astronomy, 2012, 30

\bibitem[{{Holman} \& {Murray}(2005)}]{2005Sci...307.1288H}
{Holman}, M.~J. \& {Murray}, N.~W. 2005, Science, 307, 1288

\bibitem[{{Kopal}(1950)}]{1950HarCi.454....1K}
{Kopal}, Z. 1950, Harvard College Observatory Circular, 454, 1

\bibitem[{{Lomb}(1976)}]{1976Ap&SS..39..447L}
{Lomb}, N.~R. 1976, \apss, 39, 447

\bibitem[{{Maciejewski} {et~al.}(2013){Maciejewski}, {Niedzielski},
  {Wolszczan}, {Nowak}, {Neuh{\"a}user}, {Winn}, {Deka}, {Adam{\'o}w},
  {G{\'o}recka}, {Fern{\'a}ndez}, {Aceituno}, {Ohlert}, {Errmann}, {Seeliger},
  {Dimitrov}, {Latham}, {Esquerdo}, {McKnight}, {Holman}, {Jensen}, {Kramm},
  {Pribulla}, {Raetz}, {Schmidt}, {Ginski}, {Mottola}, {Hellmich}, {Adam},
  {Gilbert}, {Mugrauer}, {Saral}, {Popov}, \& {Raetz}}]{2013AJ....146..147M}
{Maciejewski}, G., {Niedzielski}, A., {Wolszczan}, A., {et~al.} 2013, \aj, 146,
  147

\bibitem[{{Maciejewski} {et~al.}(2014){Maciejewski}, {Ohlert}, {Dimitrov},
  {Puchalski}, {Nedoroscik}, {Vanko}, {Marka}, {Baar}, {Raetz}, {Seeliger}, \&
  {Neuhauser}}]{2014AcA....64...27M}
{Maciejewski}, G., {Ohlert}, J., {Dimitrov}, D., {et~al.} 2014, \actaa, 64, 27

\bibitem[{{Mandel} \& {Agol}(2002)}]{2002ApJ...580L.171M}
{Mandel}, K. \& {Agol}, E. 2002, \apjl, 580, L171

\bibitem[{{Meschiari} {et~al.}(2009){Meschiari}, {Wolf}, {Rivera}, {Laughlin},
  {Vogt}, \& {Butler}}]{2009PASP..121.1016M}
{Meschiari}, S., {Wolf}, A.~S., {Rivera}, E., {et~al.} 2009, \pasp, 121, 1016

\bibitem[{{Mugrauer} \& {Berthold}(2010)}]{2010AN....331..449M}
{Mugrauer}, M. \& {Berthold}, T. 2010, Astronomische Nachrichten, 331, 449

\bibitem[{{M{\"u}ller} {et~al.}(2013){M{\"u}ller}, {Huber}, {Czesla}, {Wolter},
  \& {Schmitt}}]{2013A&A...560A.112M}
{M{\"u}ller}, H.~M., {Huber}, K.~F., {Czesla}, S., {Wolter}, U., \& {Schmitt},
  J.~H.~M.~M. 2013, \aap, 560, A112

\bibitem[{{Podlewska} \& {Szuszkiewicz}(2009)}]{2009MNRAS.397.1995P}
{Podlewska}, E. \& {Szuszkiewicz}, E. 2009, \mnras, 397, 1995

\bibitem[{{Rasio} \& {Ford}(1996)}]{1996Sci...274..954R}
{Rasio}, F.~A. \& {Ford}, E.~B. 1996, Science, 274, 954

\bibitem[{{Scargle}(1982)}]{1982ApJ...263..835S}
{Scargle}, J.~D. 1982, \apj, 263, 835

\bibitem[{{Steffen} {et~al.}(2012){Steffen}, {Ragozzine}, {Fabrycky}, {Carter},
  {Ford}, {Holman}, {Rowe}, {Welsh}, {Borucki}, {Boss}, {Ciardi}, \&
  {Quinn}}]{2012PNAS..109.7982S}
{Steffen}, J.~H., {Ragozzine}, D., {Fabrycky}, D.~C., {et~al.} 2012,
  Proceedings of the National Academy of Science, 109, 7982

\bibitem[{{Szab{\'o}} {et~al.}(2013){Szab{\'o}}, {Szab{\'o}}, {D{\'a}lya},
  {Simon}, {Hodos{\'a}n}, \& {Kiss}}]{2013A&A...553A..17S}
{Szab{\'o}}, R., {Szab{\'o}}, G.~M., {D{\'a}lya}, G., {et~al.} 2013, \aap, 553,
  A17

\bibitem[{{von Essen} {et~al.}(2013){von Essen}, {Schr{\"o}ter}, {Agol}, \&
  {Schmitt}}]{2013A&A...555A..92V}
{von Essen}, C., {Schr{\"o}ter}, S., {Agol}, E., \& {Schmitt}, J.~H.~M.~M.
  2013, \aap, 555, A92

\bibitem[{{Weidenschilling} \& {Marzari}(1996)}]{1996Natur.384..619W}
{Weidenschilling}, S.~J. \& {Marzari}, F. 1996, \nat, 384, 619

\end{thebibliography}

\end{document}